\documentclass[a4paper,11pt]{article}
\usepackage{pos}

\title{Towards establishing the second $b$-flavored CKM unitarity triangle}

\author*{Zhi-zhong Xing}
\author[1]{Di Zhang}

\affiliation{Institute of High Energy Physics and School of Physical Sciences, University of Chinese Academy of Sciences,\\
Beijing 100049, China}

\emailAdd{xingzz@ihep.ac.cn}
\emailAdd{zhangdi@ihep.ac.cn}
\note{Corresponding author}

\abstract{
Some fine differences between the twin $b$-flavored unitarity triangles
are calculated by means of a generalized Wolfenstein parametrization of the CKM matrix,
and a possibility of experimentally establishing the second triangle is 
briefly discussed. We find that the apexes of these two triangles, characterized 
respectively by $(\overline{\rho}, \overline{\eta})$ and $(\widetilde{\rho},
\widetilde{\eta})$, are located on the same circular arc in the complex
plane. This observation provides us with a new way to test consistency of the
CKM picture of CP violation in the quark sector and probe possible new physics. 
The differences between the apexes (i.e., $\widetilde{\rho} - \overline{\rho}$
and $\widetilde{\eta} - \overline{\eta}$) are found to be of ${\cal O}(\lambda^2)$ 
with $\lambda \simeq 0.22$ being the Wolfenstein expansion parameter, 
and the shapes of these two triangles are found to be insensitive to the
two-loop renormalization-group-equation running effects up to the accuracy
of ${\cal O}\left(\lambda^4\right)$.}

\FullConference{%
  40th International Conference on High Energy physics - ICHEP2020\\
  July 28 - August 6, 2020\\
  Prague, Czech Republic (virtual meeting)
}

\begin{document}
\maketitle

\section{Introduction}

In the standard model (SM) the quark fields interact with both the
gauge fields and the Higgs field, leading to a nontrivial mismatch between 
the flavor and mass eigenstates of quarks. This kind of mismatch, 
which is described by the $3\times 3$ Cabibbo-Kobayashi-Maskawa (CKM) matrix
$V$~\cite{Cabibbo:1963yz,Kobayashi:1973fv}, implies the existence of
flavor mixing and CP violation. The unitarity of $V$ is
strictly guaranteed by the SM itself, and it can be geometrically described 
by six unitarity triangles in the complex plane. In the past twenty years, 
the $b$-flavored unitarity triangle $\triangle^{}_s$, defined by the orthogonality 
relation $V^{*}_{ub} V^{}_{ud} + V^{*}_{cb} V^{}_{cd} + V^{*}_{tb} V^{}_{td} = 0$,
has been extensively studied. In fact, this triangle has played a significant 
role in verifying the success of the Kobayashi-Maskawa mechanism of CP violation
~\cite{Kobayashi:1973fv}. On the contrary, the other $b$-flavored triangle
$\triangle^{}_c$ (defined by $V^{*}_{ud} V^{}_{td} + V^{*}_{us}
V^{}_{ts} + V^{*}_{ub} V^{}_{tb} = 0$ in the complex plane)
has largely been ignored. The rescaled versions of $\triangle^{}_s$
and $\triangle^{}_c$, denoted respectively as
\begin{eqnarray}
\triangle^{\prime}_s ~:  1 + \frac{V^{*}_{ub} V^{}_{ud}}{V^{*}_{cb} V^{}_{cd}}
+ \frac{V^{*}_{tb} V^{}_{td}}{V^{*}_{cb} V^{}_{cd}} = 0 \; , \quad\quad
\triangle^{\prime}_c ~:  1 + \frac{V^{*}_{ud} V^{}_{td}}{V^{*}_{us} V^{}_{ts}}
+ \frac{V^{*}_{ub} V^{}_{tb}}{V^{*}_{us} V^{}_{ts}} = 0 \; ,
\end{eqnarray}
are highly similar in shape, as illustrated in Figure~1. Hence they are 
referred to as the twin $b$-flavored unitarity triangles~\cite{Xing:2019tsn}. 
The apexes of triangles $\triangle^\prime_s$ and $\triangle^\prime_c$ are
defined as 
\begin{eqnarray}
\overline{\rho} + {\rm i} \overline{\eta} =
-\frac{V^{*}_{ub} V^{}_{ud}}{V^{*}_{cb} V^{}_{cd}} \;, \quad\quad
\widetilde{\rho} + {\rm i} \widetilde{\eta} =
-\frac{V^{*}_{ub} V^{}_{tb}}{V^{*}_{us} V^{}_{ts}} \;
\end{eqnarray}
in the complex plane, respectively. An intriguing question is whether these 
two twin triangles can be distinguished by the upcoming higher-precision 
measurements to be carried out at the super-$B$ factory~\cite{Kou:2018nap} and
the high-luminosity Large Hadron Collider (LHC)~\cite{Bona:2007qt,Bediaga:2018lhg}.
\begin{figure}[h]
	\centering
	\includegraphics[width = 0.95\linewidth]{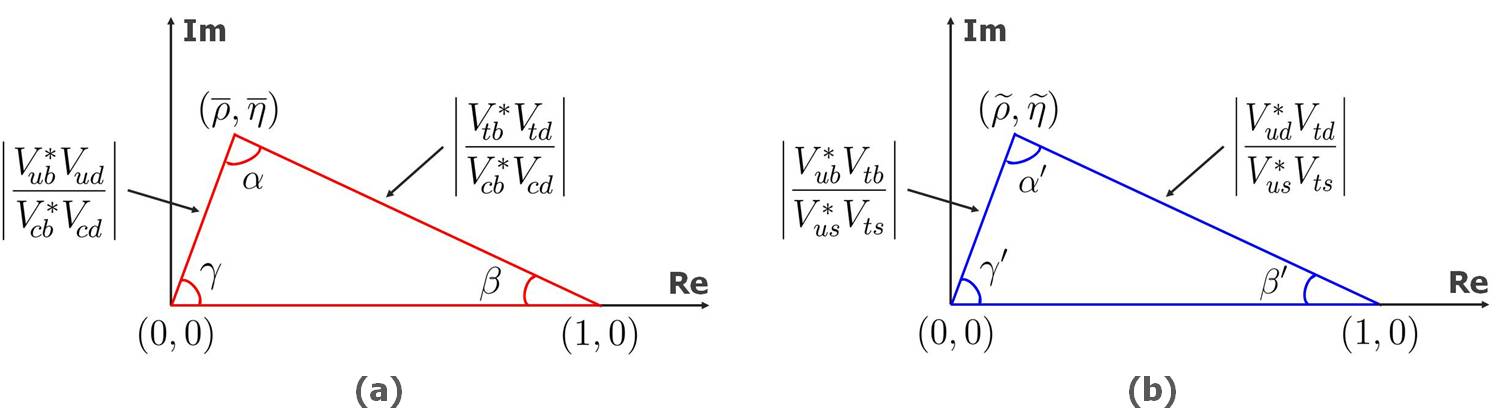}
    \vspace{0cm}
	\caption{An illustration of the twin $b$-flavored unitarity triangles 
$\triangle^\prime_s$ and $\triangle^\prime_c$ in the complex plane, where
$\left( \overline{\rho}, \overline{\eta} \right)$ and $\left( \widetilde{\rho}, 
\widetilde{\eta} \right)$ denote their apexes, respectively.}
\end{figure}

In this talk we report our recent study of fine differences between triangles
$\triangle^\prime_s$ and $\triangle^\prime_c$~\cite{Xing:2019tsn}. We find
that their apexes are actually located on a
circular arc in the complex plane. To experimentally distinguish
$\triangle^\prime_s$ and $\triangle^\prime_c$ at the $3\sigma$ level,
the apexes should be measured to the precision of $\lesssim 0.4\%$. A
possible way to separately establish $\triangle^\prime_s$ and $\triangle^\prime_c$
is to use the experimental data from $B^{\pm}_u$ and $B^0_d$-$\bar{B}^0_d$
systems and those from $B^{\pm}_u$ and $B^0_s$-$\bar{B}^0_s$ systems,
respectively. In addition, we find that the apexes and inner angles of 
$\triangle^{\prime}_s$ and $\triangle^{\prime}_c$ are insensitive to the two-loop
renormalization-group-equation (RGE) evolution up to the accuracy of $\mathcal{O}\left(\lambda^4\right)$ with $\lambda \simeq 0.22$.

\section{The apexes of $\triangle^\prime_s$ and $\triangle^\prime_c$
on a circular arc}

A popular extension of the original Wolfenstein parametrization~\cite{Wolfenstein:1983yz} 
of the CKM matrix $V$ is~\cite{Buras:1994ec,Charles:2004jd}
\begin{eqnarray}
V &\simeq& \left(\begin{matrix}1 - \frac{1}{2}\lambda - \frac{1}{8}\lambda^4 &
\lambda & A\lambda^3 \left(\rho - {\rm i}\eta\right) \\ \vspace{-0.4cm} \\ 
-\lambda + \frac{1}{2}
A^2 \lambda^5 \left[1 - 2\left(\rho + {\rm i} \eta\right)\right] & 1 - \frac{1}{2}
\lambda^2 - \frac{1}{8} \lambda^4 \left(1 + 4 A^2\right) & A \lambda^2 \\ \vspace{-0.4cm} \\ 
A\lambda^3 \left(1 -\rho - {\rm i}\eta\right) + \frac{1}{2} A\lambda^5 \left(\rho + 
{\rm i}\eta \right) & -A\lambda^2 + \frac{1}{2} A\lambda^4 \left[1 - 2\left(\rho +
{\rm i}\eta\right)\right] & 1 - \frac{1}{2} A^2\lambda^4  \end{matrix}\right)
\end{eqnarray}
up to the accuracy of $\mathcal{O}\left(\lambda^6\right)$. Note that here
$V^{}_{ub} \equiv A \lambda^3 \left(\rho - {\rm i}\eta\right)$ is {\it exact} by
definition. Moreover, the {\it exact} relationship between $(\rho, \eta)$ and
$(\widetilde{\rho} , \widetilde{\eta})$ or $(\overline{\rho} , \overline{\eta})$
is given by
\begin{eqnarray}
\rho + {\rm i} \eta = \frac{ \sqrt{1-A^2\lambda^4} \left(\overline{\rho} +
{\rm i} \overline{\eta}\right)}{\sqrt{1-\lambda^2} \left[ 1- A^2\lambda^4
\left(\overline{\rho} + {\rm i} \overline{\eta}\right)\right]} 
= \frac{ \sqrt{1-\lambda^2} \left(\widetilde{\rho} +
{\rm i} \widetilde{\eta}\right)}{\sqrt{1-A^2\lambda^4} \left[ 1- \lambda^2
\left(\widetilde{\rho} + {\rm i} \widetilde{\eta}\right)\right]} \; .
\end{eqnarray}
If both the apexes $(\widetilde{\rho} , \widetilde{\eta})$ and $(\overline{\rho} ,
\overline{\eta})$ can be directly determined from some precision measurements,
a comparison between the results of $(\rho, \eta)$ obtained independently from
Eq.~(4) will provide a meaningful consistency check
of the CKM picture for CP violation described by $\triangle^\prime_s$ and
$\triangle^\prime_c$ in the SM. Unfortunately, so far no effort has been
made towards establishing $\triangle^\prime_c$ from the available experimental
data. It is well known that the apex of $\triangle^\prime_s$ has been
excessively constrained by current experimental results for $|V^{}_{ub}|/|V^{}_{cb}|$,
$\sin 2\beta$ (CP violation in $B^0_d$ vs $\bar{B}^0_d \to J/\psi K^{}_{\rm S}$
decays), $\Delta m^{}_d$ (the mass difference of $B^0_d$-$\bar{B}^0_d$ mixing),
$\Delta m^{}_s$ (the mass difference of $B^0_s$-$\bar{B}^0_s$ mixing), $\varepsilon^{}_K$
(CP violation in $K^0$-$\bar{K}^0$ mixing) and so on~\cite{Charles:2004jd,Tanabashi:2018oca,Bona:2005vz}.
To separately constrain the apexes of $\triangle^\prime_s$ and $\triangle^\prime_c$,
one may make use of the experimental data from $B^{\pm}_u$ and $B^0_d$-$\bar{B}^0_d$
systems and those from $B^{\pm}_u$ and $B^0_s$-$\bar{B}^0_s$ systems, respectively.
The measurements of $\Delta m^{}_d$ and $\Delta m^{}_s$, which depend respectively
on $V^*_{tb} V^{}_{td}$ and $V^*_{tb} V^{}_{ts}$ via the $t$-dominated box diagrams,
are expected to be useful to distinguish between the apexes of $\triangle^\prime_s$
and $\triangle^\prime_c$. It is also possible to determine $|V^*_{tb} V^{}_{ts}|$
from more precise measurements of $B \to X^{}_s \gamma$ and $B^{}_s \to \mu^+\mu^-$
decays in the near future~\cite{Tanabashi:2018oca}.

Therefore, in the present case, we just calculate $(\rho, \eta)$ and then
$(\widetilde{\rho} , \widetilde{\eta})$ by means of of Eq.~(4) with
$\lambda = 0.22453 \pm 0.00044$, $A=0.836 \pm 0.015$, $\overline{\rho} = 0.122^{+0.018}_{-0.017}$
and $\overline{\eta}  = 0.355^{+0.012}_{-0.011}$~\cite{Tanabashi:2018oca}.
We obtain $\rho = 0.125 \pm 0.018$ and $\eta = 0.364 \pm 0.012$, together with
$\widetilde{\rho} = 0.134 \pm 0.018$ and $\widetilde{\eta} = 0.368 \pm 0.012$.
Taking advantage of Eqs.~(2) and (3), we directly arrive at
\begin{eqnarray}
\widetilde{\rho} - \overline{\rho} \simeq \left[\rho \left( 1 - \rho\right) +
\eta^2 \right] \lambda^2 \; , \quad\quad \widetilde{\eta} - \overline{\eta} \simeq
\eta \left( 1 - 2\rho \right) \lambda^2 \; ,
\end{eqnarray}
up to the accuracy of ${\cal O}(\lambda^4)$. So the differences
$\widetilde{\rho} - \overline{\rho}$ and $\widetilde{\eta} - \overline{\eta}$
remain within the error bars of these four parameters. We expect that $\triangle^\prime_c$
and $\triangle^\prime_s$ will be experimentally distinguishable at the $3\sigma$
level if their apexes $(\overline{\rho} , \overline{\eta})$ and
$(\widetilde{\rho} , \widetilde{\eta})$ can be determined to the precision of 
$\lesssim 0.4\%$.

Let us proceed to take a look at the inner angles of triangles $\triangle^\prime_s$
and $\triangle^\prime_c$, which are defined as
\begin{eqnarray}
\alpha  &\equiv&  \arg \left( - \frac{V^{*}_{tb} V^{}_{td}}{V^{*}_{ub}V^{}_{ud}}
\right) \;,\quad\quad  
\beta  \equiv  \arg \left(-\frac{V^{*}_{cb}V^{}_{cd}}{V^{*}_{tb}V^{}_{td}}
\right) \;, \quad\quad \gamma  \equiv  \arg \left( - \frac{V^{*}_{ub}V^{}_{ud}}{V^{*}_{cb}V^{}_{cd}} \right) \;,
\nonumber \\
\alpha^\prime & \equiv & \arg \left( - \frac{V^{*}_{ud} V^{}_{td}}{V^{*}_{ub} V^{}_{tb}}
\right) \;, \quad\quad \beta^\prime  \equiv  \arg \left( - \frac{V^{*}_{us}V^{}_{ts}}
{V^{*}_{ud}V^{}_{td}} \right) \;, \quad\quad \gamma^\prime  \equiv  \arg \left( -
\frac{V^{*}_{ub}V^{}_{tb}}{V^{*}_{us}V^{}_{ts}} \right) \;.
\end{eqnarray}
With the help of Eqs.~(3) and (6), one can achieve that $\alpha^\prime = \alpha$
holds exactly by definition, and
\begin{eqnarray}
\beta^\prime - \beta = \gamma - \gamma^\prime \simeq \eta \lambda^2 \left[ 1 +
\left(\frac{1}{2} -A^2 -\rho\right) \lambda^2\right] \; .
\end{eqnarray}
Taking account of the values of $\lambda$, $A$, $\overline{\rho}$ and $\overline{\eta}$,
we immediately obtain $\alpha = \alpha^\prime \simeq 87.0^\circ \pm 2.5^\circ$,
$\beta \simeq 22.0^\circ \pm 0.8^\circ$, $\gamma \simeq 71.0^\circ \pm 2.6^\circ$,
$\beta^\prime \simeq 23.0^\circ \pm 0.8^\circ$ and $\gamma^\prime \simeq 70.0^\circ
\pm 2.6^\circ$. So the numerical results of $\beta^\prime - \beta$ and $\gamma -
\gamma^\prime$, which also characterize the tiny difference between triangles
$\triangle^\prime_s$ and $\triangle^\prime_c$, remain within the error bars of these
four angles and hence require more accurate measurements.
\begin{figure}[t]
\centering
\includegraphics[width = 0.95\linewidth]{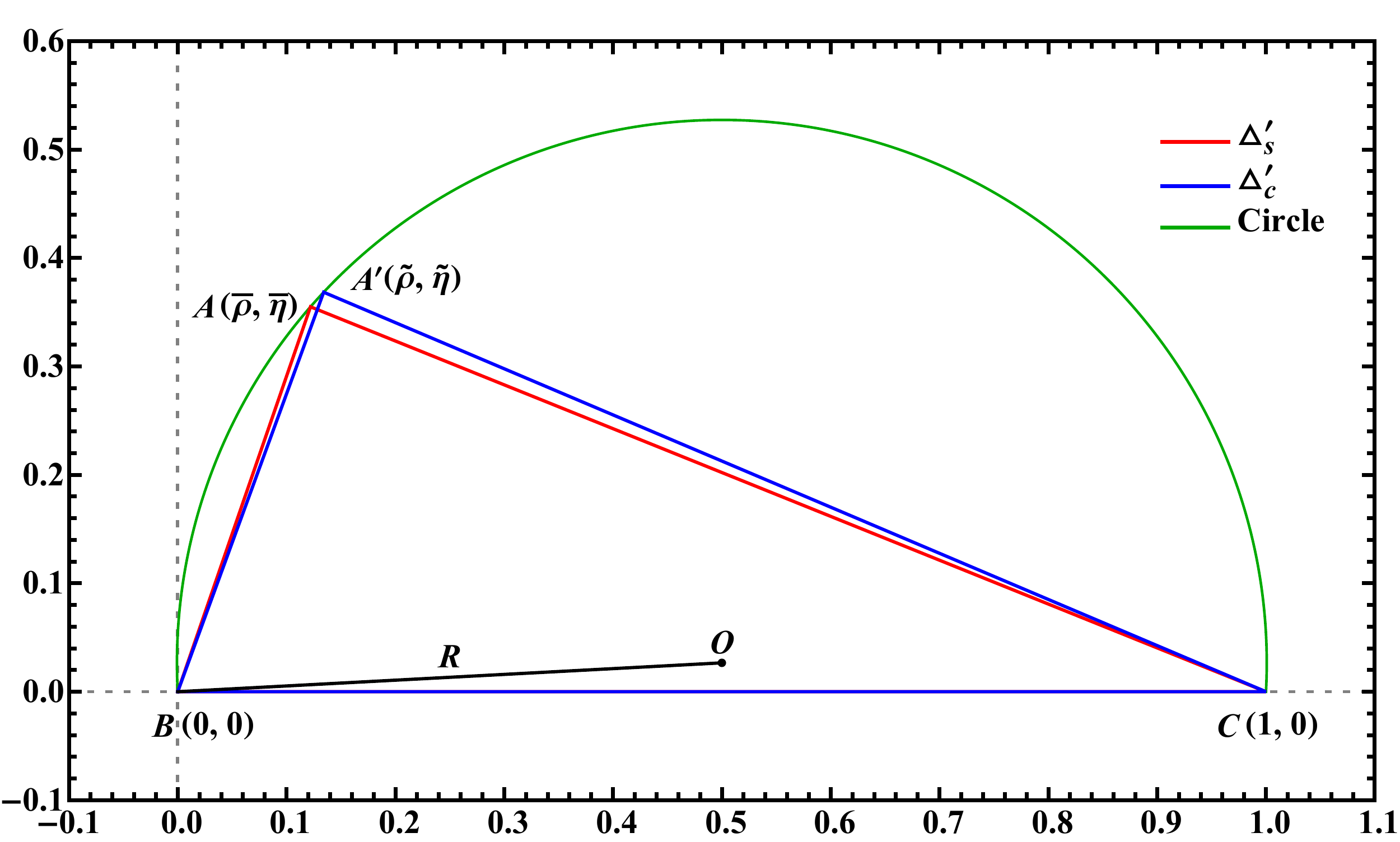}
\caption{The rescaled CKM unitarity triangles $\triangle^\prime_{s} = \triangle ABC$
and $\triangle^\prime_{c} = \triangle A^\prime BC$ in the complex plane, where 
the center and radius of the
circular arc are $O = (0.5, 0.5 \cot\alpha)$ and $R = 0.5 \csc\alpha$.}
\end{figure}

Now that the twin rescaled unitarity triangles $\triangle^\prime_c$ and
$\triangle^\prime_s$ share a common inner angle $\alpha^\prime = \alpha$
and a common side $BC$ as shown in Figure~2, their corresponding apexes
$(\widetilde{\rho},\widetilde{\eta})$ and $(\overline{\rho},\overline{\eta})$
must be located on a circular arc in the upper complex plane. That is
\begin{eqnarray}
\left(\widetilde{\rho} - \frac{1}{2}\right)^2 + \left(\widetilde{\eta} -
\frac{1}{2}\cot{\alpha}\right)^2 = \left(\overline{\rho} - \frac{1}{2}\right)^2 +
\left(\overline{\eta} - \frac{1}{2}\cot{\alpha}\right)^2 = \left(\frac{1}{2}
\csc\alpha \right)^2 \;.
\end{eqnarray}
It is obvious that the center and radius of the circular arc determined by Eq.~(8)
and shown in Figure~2 are $O = \left(0.5, 0.5\cot\alpha\right)$ and $R = 0.5\csc\alpha$.
The fact that all the apexes of $\triangle^\prime_c$ and $\triangle^\prime_s$
are located on the same circular arc is of course a natural consequence of
the CKM unitarity. It provides another interesting way to test
the consistency of quark flavor mixing and CP violation in the SM. Since
$\triangle^\prime_s$ has been established to a very good degree of accuracy,
it allows us to fix a benchmark circular arc as shown in Figure~2. The future
measurements of $(\widetilde{\rho},\widetilde{\eta})$ will tell us to what
extent the experimental values of this apex are also located on the same
circular arc. In other words, an experimental test of the equality given in
Eq.~(8) will be greatly useful at both the super-$B$ factory and the
high-luminosity LHC.

\section{Two-loop RGE evolution of $\triangle^\prime_s$ and $\triangle^\prime_c$}

Note that elements of the CKM matrix $V$ depend on the energy scale $\Lambda$.
When $\Lambda$ is far above the electroweak scale $\Lambda^{}_{\rm EW} \sim
10^2~{\rm GeV}$, the RGE running effects of $V$ will become appreciable and
should be taken into account.
In particular, the two-loop RGEs of $V$ have been derived in the framework of
the SM or its minimal supersymmetric version (MSSM)~\cite{Machacek:1983fi,Barger:1992ac,Barger:1992pk,Luo:2002ey}.
In view of $y^{}_u/y^{}_c \sim y^{}_c/y^{}_t \sim \lambda^4$, $y^{}_d/y^{}_s
\sim y^{}_s/y^{}_b \sim \lambda^2$ at a given energy scale 
and the relatively strong hierarchies of those off-diagonal elements of $V$,
Barger {\it et al} have found~\cite{Barger:1992pk}
\begin{eqnarray}
\frac{\rm d}{{\rm d}t} \left( \begin{matrix} |V^{}_{ud}| & |V^{}_{us}| & |V^{}_{ub}|
\\ |V^{}_{cd}| & |V^{}_{cs}| & |V^{}_{cb}| \\ |V^{}_{td}| & |V^{}_{ts}| & |V^{}_{tb}|
\end{matrix}\right) \simeq \left(S^{}_{1} + S^{}_{2} \right)  \left(
\begin{matrix} 0 & 0 & |V^{}_{ub}| \\ 0 & 0 & |V^{}_{cb}| \\ |V^{}_{td}| & |V^{}_{ts}|
& 0 \end{matrix}\right) \; ,\quad\quad  \frac{{\rm d} {\cal J}}{{\rm d}t} \simeq 2
\left(S^{}_{1} + S^{}_{2} \right) \mathcal{J} \; ,
\end{eqnarray}
where $t \equiv \ln \left(\Lambda/\Lambda^{}_{\rm EW}\right)$, ${\cal J} \equiv
{\rm Im} \left(V^{}_{ud} V^{}_{cs} V^*_{us} V^*_{cd}\right) \simeq A^2 \lambda^6 \eta$ 
is the Jarlskog invariant of CP violation~\cite{Jarlskog:1985ht}, 
$S^{}_{1}$ and $S^{}_{2}$ stand respectively
for the one- and two-loop contributions to the RGEs of $V$ (and their explicit 
definitions can be found in Ref.~\cite{Xing:2019tsn}).

After a careful check of the approximations made in obtaining Eq.~(9), we conclude
that the two-loop RGEs shown in Eq.~(9) are valid up to the accuracy
of ${\cal O}(\lambda^4)$. To the same order, the integral solutions of the
Wolfenstein parameters to Eq.~(9) can be figured out as follows:
\begin{eqnarray}
&& \lambda(\Lambda) \simeq \lambda(\Lambda_{\rm EW}) \;, \quad\quad  \rho(\Lambda)
\simeq \rho(\Lambda_{\rm EW}) \;, \quad\quad  \eta(\Lambda) \simeq \eta(\Lambda_{\rm EW}) \;,
\quad\quad  A(\Lambda) \simeq I^{}_{1} I^{}_{2} A(\Lambda_{\rm EW}) \; ; \hspace{0.6cm}
\nonumber \\
&& \overline{\rho}(\Lambda) \simeq \overline{\rho}(\Lambda_{\rm EW}) \;, \quad\quad
\overline{\eta}(\Lambda) \simeq \overline{\eta}(\Lambda_{\rm EW}) \;, \quad\quad
\widetilde{\rho}(\Lambda) \simeq \widetilde{\rho}(\Lambda_{\rm EW}) \;, \quad\quad
\widetilde{\eta}(\Lambda) \simeq \widetilde{\eta}(\Lambda_{\rm EW}) \;,
\end{eqnarray}
where $\Lambda$ denotes an arbitrary energy scale between $\Lambda^{}_{\rm EW}$
and $\Lambda^{}_{\rm GUT}$, and the loop functions $I^{}_1$ and $I^{}_2$ have also been
defined in Ref.~\cite{Xing:2019tsn}. It is clear that $|V^{}_{ub}|$,
$|V^{}_{cb}|$, $|V^{}_{td}|$ and $|V^{}_{ts}|$ evolve in the same way as $A$,
and $\mathcal{J}(\Lambda) \simeq I^2_{1} I^2_{2} \mathcal{J}(\Lambda_{\rm EW})$
holds for the Jarlskog invariant. In comparison, $|V^{}_{ud}|$, $|V^{}_{us}|$,
$|V^{}_{cd}|$, $|V^{}_{cs}|$ and $|V^{}_{tb}|$ are essentially stable against
changes of the energy scale $\Lambda$.
Thus the rescaled unitarity triangles $\triangle^\prime_s$ and $\triangle^\prime_c$
keep unchanged when the energy scale $\Lambda$ evolves from $\Lambda^{}_{\rm EW}$
to $\Lambda^{}_{\rm GUT}$ or vice versa, up to the
accuracy of ${\cal O}(\lambda^4)$. In other words, the overall shape of either
of $\triangle^{}_s$ and $\triangle^{}_c$ keeps undeformed up to the same accuracy.

In summary, we have discussed whether the twin $b$-flavored unitarity
triangles $\triangle^\prime_s$ and $\triangle^\prime_c$ can be experimentally
distinguished from each other in the context of the upcoming precision measurements 
at the super-$B$ factory and the high-luminosity LHC. The answer is affirmative if
their apexes are measured to a sufficiently good degree of accuracy. 
We have pointed out that the apexes $(\overline{\rho}, \overline{\eta})$ and
$(\widetilde{\rho}, \widetilde{\eta})$ are exactly located on a circular arc,
whose center and radius are $O = (0.5, 0.5 \cot\alpha)$ and $R = 0.5 \csc\alpha$
respectively. We have also shown that $(\overline{\rho}, \overline{\eta})$ and
$(\widetilde{\rho}, \widetilde{\eta})$ are insensitive to the two-loop RGE
running effects up to the accuracy of ${\cal O}(\lambda^4)$. 
So the experimental results of all
the inner angles of $\triangle^\prime_s$ and $\triangle^\prime_c$ obtained
at low energies can directly be confronted with some theoretical predictions
at a superhigh energy scale.

This work was supported in part by the National Natural Science Foundation of
China under grant No. 11775231, grant No. 11835013 and grant No. 12075254.

\end{document}